\documentclass[twocolumn,aps,prd,superscriptaddress,amsmath,amssymb]{revtex4-1}
\usepackage{graphicx}% Include figure files

\usepackage[latin1]{inputenc}
\usepackage[english]{babel}
\usepackage[T1]{fontenc}
\usepackage{hyperref}
\usepackage{multirow}
\usepackage{sistyle}

\renewcommand{\d}{\text{d}}
\newcommand{\fref}[1]{Fig. \ref{#1}}
\newcommand{\eref}[1]{Eq. ~(\ref{#1})}

\begin{document}
\title{High energy cosmic-ray interactions with particles from the Sun}
\date{\today}
\author{Kristoffer K. Andersen}
%\email[E-mail:]{kka@phys.au.dk}
\affiliation{Department of Physics and Astronomy, Aarhus University, DK-8000 Aarhus C, Denmark}
\affiliation{Lawrence Berkeley National Laboratory, Berkeley CA, 94720 USA}
\author{Spencer R. Klein}
\affiliation{Lawrence Berkeley National Laboratory, Berkeley CA, 94720 USA}
\affiliation{Department of Physics, University of California, Berkeley, CA, 94720 USA}
%\hyphenation{}

\begin{abstract}
Cosmic-ray protons with energies above $10^{16}$ eV passing near the Sun may interact with photons emitted by the Sun and be excited to a $\Delta^+$ resonance.  When the $\Delta^+$ decays, it produces pions which further decay to muons and photons which may be detected with terrestrial detectors.  A flux of muons, photon pairs (from $\pi^0$ decay), or individual high-energy photons coming from near the Sun would be a rather striking signature, and the flux of these particles is a fairly direct measure of the flux of cosmic-ray nucleons, independent of the cosmic-ray composition.   In a solid angle within $15^\circ$ around the Sun the flux of photon pairs is about $\SI{1.3e-3}{}$ particles/(km$^2\cdot$yr), while the flux of muons is about $\SI{0.33e-3}{}$ particles/(km$^2\cdot$yr).  This is beyond the reach of current detectors like the Telescope Array, Auger, KASCADE-Grande or IceCube.  However, the muon flux might be detectable by next-generation air shower arrays or neutrino detectors such as ARIANNA or ARA.  We discuss the experimental prospects in some detail. Other cosmic-ray interactions occuring close to the Sun are also briefly discussed.

\end{abstract}

\date{\today}
\maketitle

\section{Introduction}
% Introduction
Despite decades of effort, the composition of cosmic rays at high energies is still an unsettled question.
Many balloon and satellite based experiments have measured the composition at moderate energies, up to the 'knee' of the cosmic-ray spectrum between $10^{15}$ eV and  $10^{16}$ eV \cite{PDG2010}. These experiments are not large enough to collect significant data at higher energies.  Surface-based experiments extend to higher energies.  Unfortunately, they cannot measure the fluxes of individual elements, but instead study average composition.  For energies around the knee the Tibet air-shower array \cite{Amenomori2011}, KASCADE-Grande \cite{Apel200986} and EAS-TOP \cite{Aglietta2004583} all found  a similar increase in average atomic number, $\langle A\rangle$ as the cosmic-ray energy increased.  At higher energies, between $10^{18}$ eV and $10^{19.5}$ eV, the HiRes collaboration reported that protons dominate \cite{Belz20095}.   However, the Auger collaboration reports a change in composition, from mostly protons around $10^{18}$ eV to mostly iron around $10^{19.5}$ eV \cite{PhysRevLett.104.091101}. 

The cosmic-ray flux may have galactic and extra-galactic components \cite{2compJETP}.  Lower energy cosmic rays are likely produced within our galaxy; the energy-dependence of the composition is consistent with the leaky box model \cite{Cesarsky1980} where the chance of escaping the galaxy depends on the charge of the cosmic ray.  More highly charged ions are contained longer in the galaxy compared to particles with a lower charge, since the gyromagnetic radius is inversely proportional to the cosmic-ray charge.  At higher energies, there is an apparent extra-galactic component which conventional wisdom assumes is mostly protons.   

In this article we explore a novel way of determining  the flux of cosmic-ray protons by measuring the interactions between cosmic rays and photons from the Sun.  Cosmic-rays that pass nearby the Sun may interact with photons emitted by it.  The effect of photodissociation of cosmic-ray nuclei by photons has been investigated by Laf\`{e}bre \textit{et al.} \cite{GZe2008}.  Here, we consider a different reaction, the photoexcitation of cosmic-ray protons to a $\Delta^+$ resonance.  This $\Delta^+$ decays, producing pions which themselves decay to muons or photons which may be observed at the Earth.   When the cosmic-ray protons have a high enough Lorentz boost, optical photons are energetic enough to cause this excitation.   A $\Delta^+ (1232)$ resonance may be produced when a proton with an energy of about $1.6\times10^{17}$ eV collides head-on with an optical (1 eV) photon.  A measurement of the $\approx 10^{16}$ eV photon or muon flux coming from the vicinity of the Sun would provide a fairly direct measurement of the flux of cosmic-ray nucleons at an energy around $10^{17}$ eV.   We consider the possibilities of observing single muons and photons, as well as the possibility of observing multi-core showers produced in a $\pi^0$ decay.  For heavier cosmic rays similar interactions can occur for the individual nucleons bound in the nuclei.  

Besides photoproduction of $\Delta^+$ resonances we also investigate inelastic interactions between cosmic rays and protons in the atmosphere of the Sun. For cosmic rays with energies below 1 TeV where magnetic deflection by the solar magnetic field is important these intereactions have been investigated by Seckel \textit{et al.} \cite{gaisser1991}. Here we calculate the flux of photons from the decay of pions created when high-energy cosmic rays, where magnetic deflection is negligible, interact with the solar atmosphere. Another possible interaction is for solar photons to produce an $e^+e^-$ pair in the field of a cosmic ray. This is possible since the solar photons have MeV energies in the Lorentz boosted frame of the high energy cosmic ray and produces high-energy electrons and positrons.

\section{Approach}
% delta resonances
We first calculate the interaction probability for a single cosmic ray with energy $E_{CR}$ passing near the Sun at impact parameter $b$. The flux on Earth is found by integrating over the energies given by the cosmic-ray spectrum and the observed solid angle around the Sun. For interactions leading to muons we take into account the probability of the muons decaying before reaching the Earth.

\begin{figure}[tb]
\centering
	\includegraphics[width=\columnwidth]{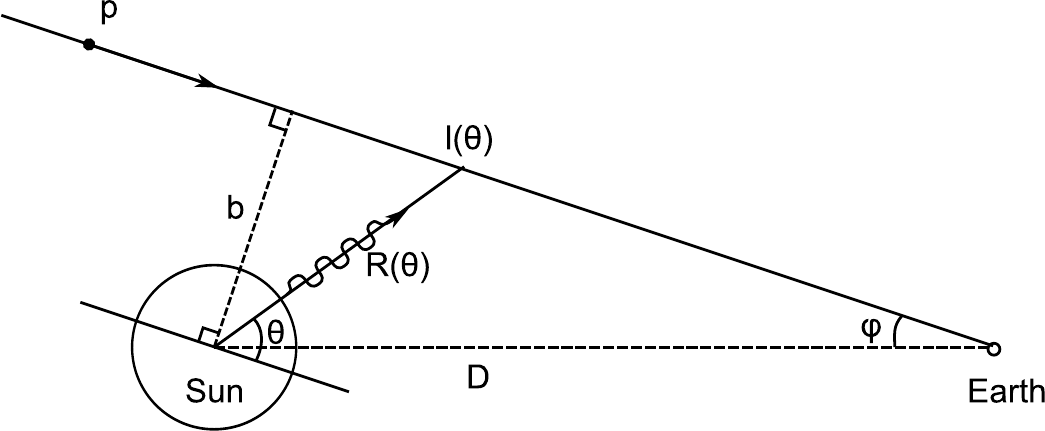}
	\caption{\label{fig:draw} The geometry of the interaction. The solar photon will interact with the cosmic ray at trajectory point $l(\theta)$. The trajectory passes by the Sun at distance (impact parameter) $b$, at an angular separation $\varphi$ from the Sun. $D$ is the distance from the Sun to the Earth.}
\end{figure}

Consider a single cosmic-ray particle travelling at an impact parameter $b$ relative to the Sun with energy $E_{CR}$ and mass $m$ as illustrated in \fref{fig:draw}. For an interaction angle $\theta$, the position of the interaction point is given by $l(\theta)$ and the distance from the center of the Sun is $R(\theta)$. The deflection of the cosmic ray in the magnetic field of the Sun is negligible, since the energy of the cosmic ray is above $10^{16}$ eV. The interaction probability is 
\begin{equation}
P(E_{CR},b)=\int \d t \int_0^{\infty} \frac{\d k}{h} \	\sigma(k,E_{CR},\theta) \times N_\gamma(k,R(\theta)),
	\label{eq:P}
\end{equation}
where $t$ is the time, $k$ is the photon energy, $h$ is Planck's constant, $N_\gamma(k,R(\theta))$ is the flux of photons with energy $k$ from the Sun at a distance $R(\theta)$ and $\sigma(k,E_{CR},\theta)$ is the interaction cross section. The integral over the transit time can conveniently be transformed to an integral over the interaction angles $\theta$. Since the particles are ultrarelativistic $ct = l(\theta) = b/\tan\theta$, where $c$ is the speed of light. With this substitution the interaction probability becomes 
\begin{eqnarray}
	P(E_{CR},b)=\int_\varphi^\pi \d\theta \frac{b}{c\sin^2\theta} \int_0^\infty \frac{\d k}{h}\times \nonumber\\*	\sigma(k,E_{CR},\theta) \times N_\gamma(k,R(\theta)),
	\label{eq:P2}
\end{eqnarray}
where $\varphi = \sin^{-1}(b/D)$ is the angular separation defined in \fref{fig:draw}.

The photon flux from the Sun is assumed to be a blackbody spectrum. The energy flux $I(k)$ is given by Planck's law. Hence, the photon flux per steradian is 
\begin{equation}
	N_\text{Pl}(k)\frac{\d k}{h} =\frac{I(k)}{k}\d\nu=2\frac{\nu^2}{ c^2}\times\frac{\d\nu}{\exp(k/k_bT)-1}, \label{eq:Ng}
\end{equation}
for a $\d\nu$ frequency interval, $k_b$ is Boltzmann's constant and $T=5778$ K is the temperature of the Sun. At distance $R(\theta)$ the photon flux is scaled by the surface area factor $R^2_\text{Sun}/R(\theta)^2$ where $R_\text{Sun}$ is the radius of the Sun. 
For a small surface area of the Sun the radiation spreads into a half sphere and since the radiation is Lambertian this gives a solid angle factor of $\pi$. A more accurate calculation would include the finer details in the solar spectrum (the actual photon spectrum is higher than the blackbody spectrum at high energies and lower around 200 nm \cite{AschwandenCorona}). Neglecting these details, the photon flux at distance $R(\theta)$ becomes 
\begin{equation}
	N_\gamma(k,R(\theta))=\pi\frac{R^2_\text{Sun}}{R(\theta)^2}N_\text{Pl}(k).
	\label{eq:NgR}
\end{equation}

\begin{figure}[tb]
	\centering
		\includegraphics[width=\columnwidth]{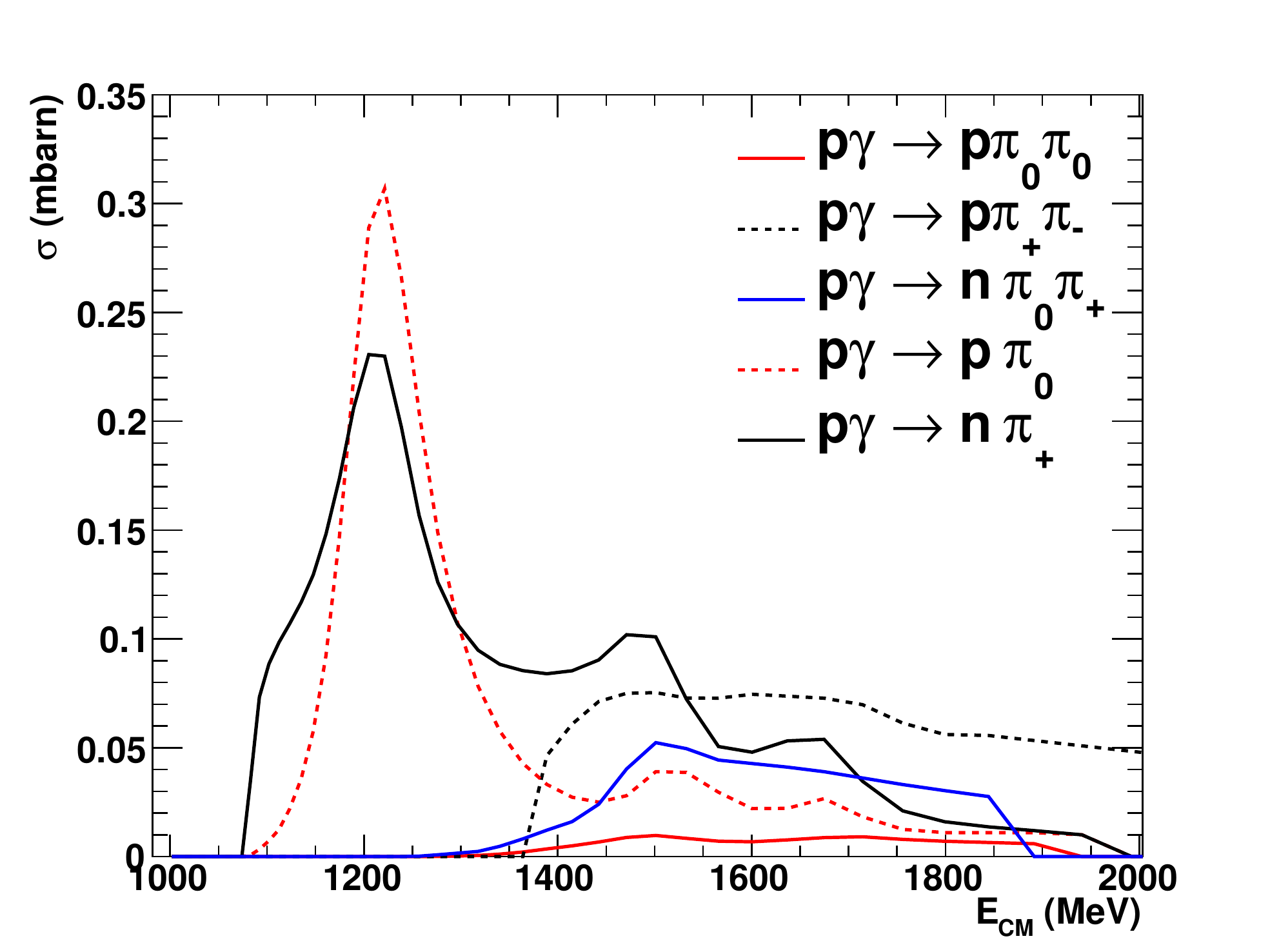}
	\caption{	\label{fig:cross}(Color online) The cross sections for photoproduction of pions. The single pion production cross section is dominated by the $\Delta^+$ resonance.}
\end{figure}

The cross sections for photoproduction of pions are well measured \cite{KlemptBaryon}. For the single pion cases; $p\gamma \rightarrow p\pi^{0}$ and $p\gamma \rightarrow n\pi^{+}$, cross sections can be found at the Center for Nuclear Studies. Cross sections from their SAID Partial Wave Analysis \cite{SAID} are used for these reactions. Their analysis show a good agreement with measured cross sections \cite{SAID2008}. For reactions that produce final states with more than one pion the cross section sources are listed in Table \ref{tab:pidata}. These cross sections are plotted in \fref{fig:cross}.   To a good approximation, this data can be fit by analytic Breit-Wigner resonances for the $\Delta(1232)$ and $\Delta(1600)$ resonances with the parameters given in the Particle Data Book \cite{PDG2010}. 
 
\begin{table}
	\caption{\label{tab:pidata} Data sources for multi-pion cross sections.}
	\begin{ruledtabular}
	\begin{tabular}{ll}
	Interaction &  Source \\ \hline
	$p\gamma \rightarrow p\pi^{+}\pi^{-}$ & Wu \textit{et al.} \cite{WU2005} \\ \hline
\multirow{2}{0.5\columnwidth}{$p\gamma \rightarrow p\pi^{0}\pi^{0}$}	 & Thoma \textit{et al.} \cite{THOMA2008} \\
																				&	Wolf \textit{et al.} \cite{WOLF2000} \\ \hline
	\multirow{2}{0.5\columnwidth}{$p\gamma \rightarrow n\pi^{+}\pi^{0}$} & Langg\"{a}rtner \textit{et al.} \cite{LANGGARTNER2001} \\
	 & Fix \textit{et al.} \cite{FIX2005}
	\end{tabular}
	\end{ruledtabular}
\end{table}

$\Delta$ production is also possible in nuclei with $Z>1$.  Photoproduction data shows that the $\Delta$ resonance is somewhat washed out as the nuclear mass rises \cite{ArmstrongNuclPB}.  However, to a fairly good approximation, one can treat the individual nucleons in a nucleus as if they are independent. In other words to a good approximation the cross section per nucleon is constant for a wide range of nuclei \cite{ChristillinPRep}. Also for heavier nuclei the $\Delta$-resonance lies at slightly higher energies \cite{CarlosNuclPA}. These differences are relatively small, so for our calculation the flux of photons and muons is sensitive to the total flux of nucleons in cosmic rays, largely irrespective of whether they are combined into heavier nuclei or not.

The flux of created particles is given by the cosmic-ray spectrum $N_{CR}$ multiplied by the interaction probability $P(E_{CR},b)$. For simplicity and since there are no conclusive measurements of the protonic cosmic-ray spectrum in the relevant energy regime, we assume that all cosmic rays are protons. In the region of interest (between the knee ($E_\text{knee} = \SI{4e15}{eV}$) and the ankle ($E_\text{ankle}\sim10^{19}\SI{}{eV}$)) we assume that the spectrum follows a power law \cite{PDG2010} 
\begin{equation}
N_{CR}(E)=\SI{3.5e24}{\frac{eV^2}{m^2.s.sr}}\times E^{-3}.
	\label{eq:Ncr}
\end{equation}

The created $\pi^+$ ($\pi^0$)  decay with lifetimes of 26 ns ($\SI{84e-9}{ns}$) producing muons and neutrinos, and two photons respectively.   We assume that created photons will reach the Earth without interacting with the interplanetary medium. However, the muon lifetime is $\tau_{\mu}=\SI{2.2}{\mu s}$ so they are likely to decay before reaching Earth. The $\mu$ survival probability is
\begin{equation}
	P_\mu(\theta,b,\gamma_\mu) = \exp \left(- \frac{b/\tan\varphi-l(\theta)}{c\tau_\mu \gamma_\mu} \right),
	\label{eq:Pmu}
\end{equation}
where $b/\tan\varphi-l(\theta)$ is the distance from the interaction point to the Earth and $\gamma_\mu$ is the Lorentz boost of the muon, which is approximately equal to the Lorentz factor of the pion. We approximate $\gamma_\mu$ by the following expression
\begin{equation}
	\gamma_\mu = \frac{E_{CM}^2-m_p^2+m_\pi^2}{2E_{CR} m_\pi}\sqrt{1+\frac{E_{CR}^2}{E_{CM}^2}}.
	\label{eq:gamMu}
\end{equation}
$E_{CM}$ is the center of mass energy of the cosmic ray and the solar photon, $m_p$ is the mass of the proton, and $m_\pi$ is the pion mass. This derivation assumes that the momentum of the pion along the direction of the cosmic ray is negligible in the center of mass frame; this is true on the average, but individual $\mu$ will have slightly higher or lower Lorentz boosts.
 
\begin{figure}[tb]
	\centering
  	\includegraphics[width=\columnwidth]{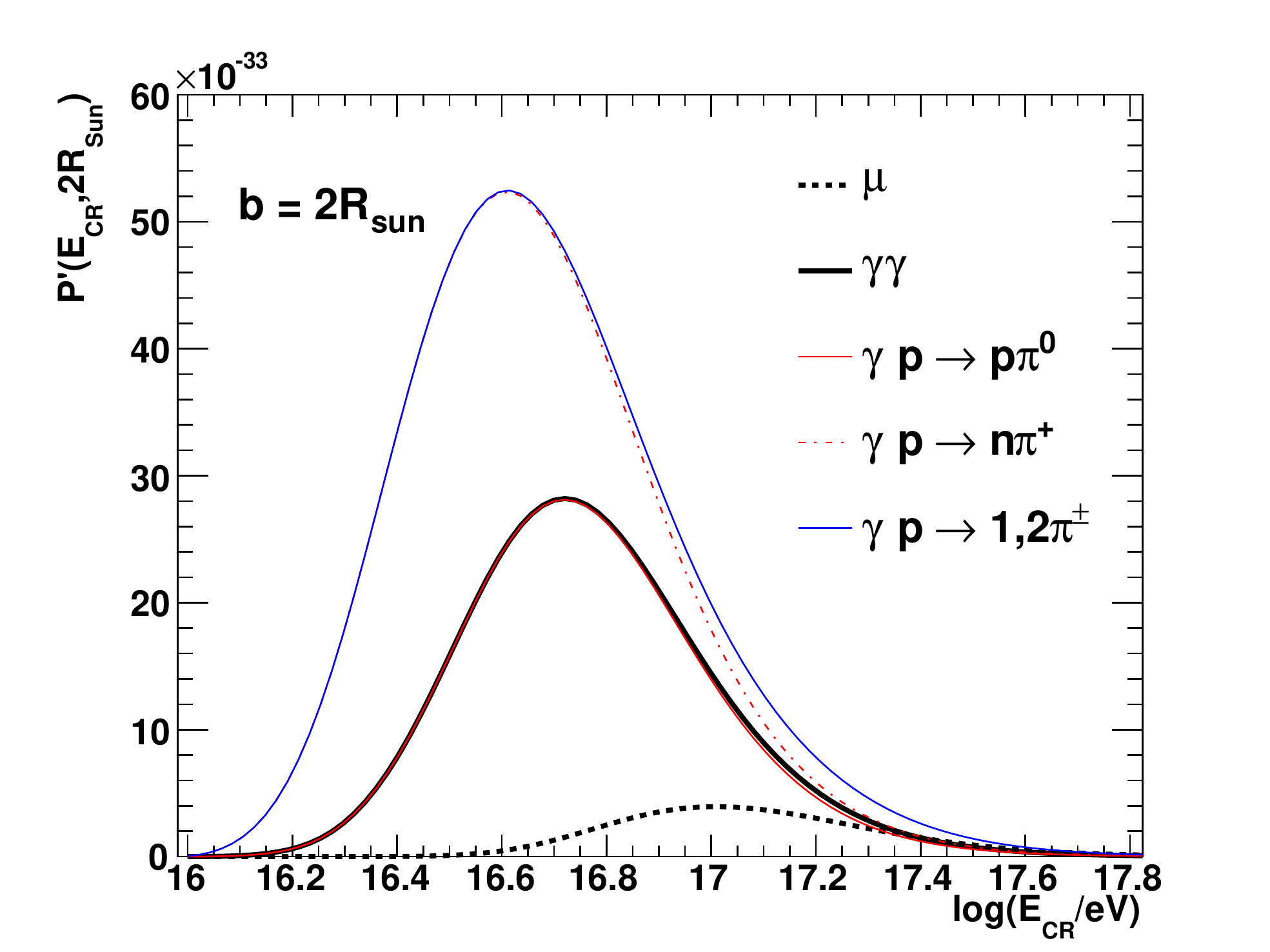}
		\caption{	\label{fig:ps}(Color online) Number of pions produced together with number of muons and photon pairs reaching Earth per cosmic ray at $b=2R_\text{Sun}$. $P'$ is equal to \eref{eq:P2} except for muons where the survival probability given by \eref{eq:Pmu} is included. Thick lines show photon pairs (solid) and muons (broken). Thin lines show the total flux of charged pions (blue), flux of single charged pion (red dash-dot) and single neutral pion (red solid) emission. The total flux of neutral pions is equal to the flux of photon pairs since the survival probability is one.}
\end{figure}

For two-pion emission it is more difficult to calculate the survival probability, because the kinematics are more complicated.  In these calculations it is assumed that all two-pion decays are sequential with a $\Delta(1232)$ resonance as intermediate state, and that the resonance width may be neglected. All pions are assumed to be emitted perpendicular to the direction of the decaying resonance. This is a simplified model, but since interactions where two pions are produced occur at higher $E_{CM}$ they contribute little compared to single pions as seen in \fref{fig:ps}. 

Including the survival probability for muons and assuming that all photons from the $\pi^0$ decay reach Earth we can calculate the number of particles reaching Earth per cosmic ray at impact parameter $b$. This is shown in \fref{fig:ps}. The low energy muons decay since the Lorentz factor for these particles is too small, and therefore the lifetime is not sufficiently Lorentz boosted. This may effect the calculations of the muon flux significantly, since the calculation of the survival probability for muons is relatively simple for two pion emission, and these are more important at high cosmic ray energies. The muons decay to positrons and neutrinos. Air showers from these positrons will be indistinguishable to photon initiated showers.

The flux of particles reaching the Earth is obtained by integrating over the observed solid angle around the Sun. The interaction probability depends only on the cosmic-ray energy and the impact parameter. One can therefore conveniently integrate over impact parameter and use the relationship between the observation angle $\varphi$ and the impact parameter, $\sin\varphi=b/D$ where $D$ is the distance from the Sun to the Earth. The flux is therefore

\begin{eqnarray}
	\Phi = 2\pi \int_{R_\text{Sun}}^{b_\text{max}}\frac{b\d b}{D\sqrt{D^2-b^2}} \times \notag\\
	\int \d E_{CR} \ N_{CR}(E_{CR})P'(E_{CR},b)
	\label{eq:phi}
\end{eqnarray}
where $P'(E_{CR},b)$ is equal to $P(E_{CR},b)$ given by \eref{eq:P2} but includes the survival probability for muons given by \eref{eq:Pmu}. $b_\text{max} = D\sin\varphi_\text{max}$ is given by the maximum observation angle $\varphi_\text{max}$.

\begin{figure}[tb]
	\centering
		\includegraphics[width=\columnwidth]{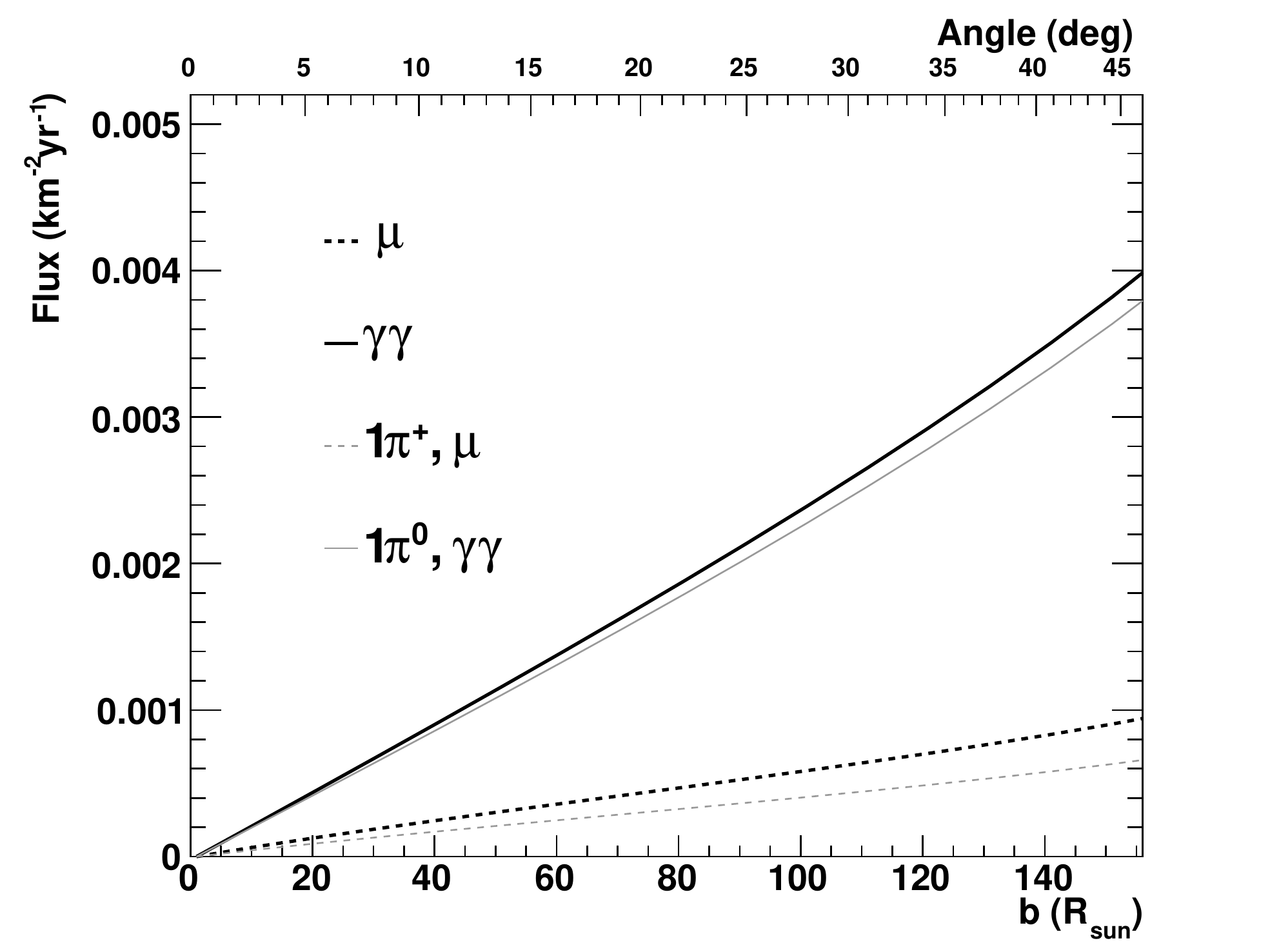}
		\caption{	\label{fig:tot}Flux of muons (black broken line) and photon (black solid line) pairs on Earth coming from within an impact parameter $b$ corresponding to the angle $\varphi = \sin^{-1}(b/D)$. The gray lines indicate the flux when neglecting processes leading to 2 pion states.}
\end{figure}

\begin{figure}[tb]
	\centering
		\includegraphics[width=\columnwidth]{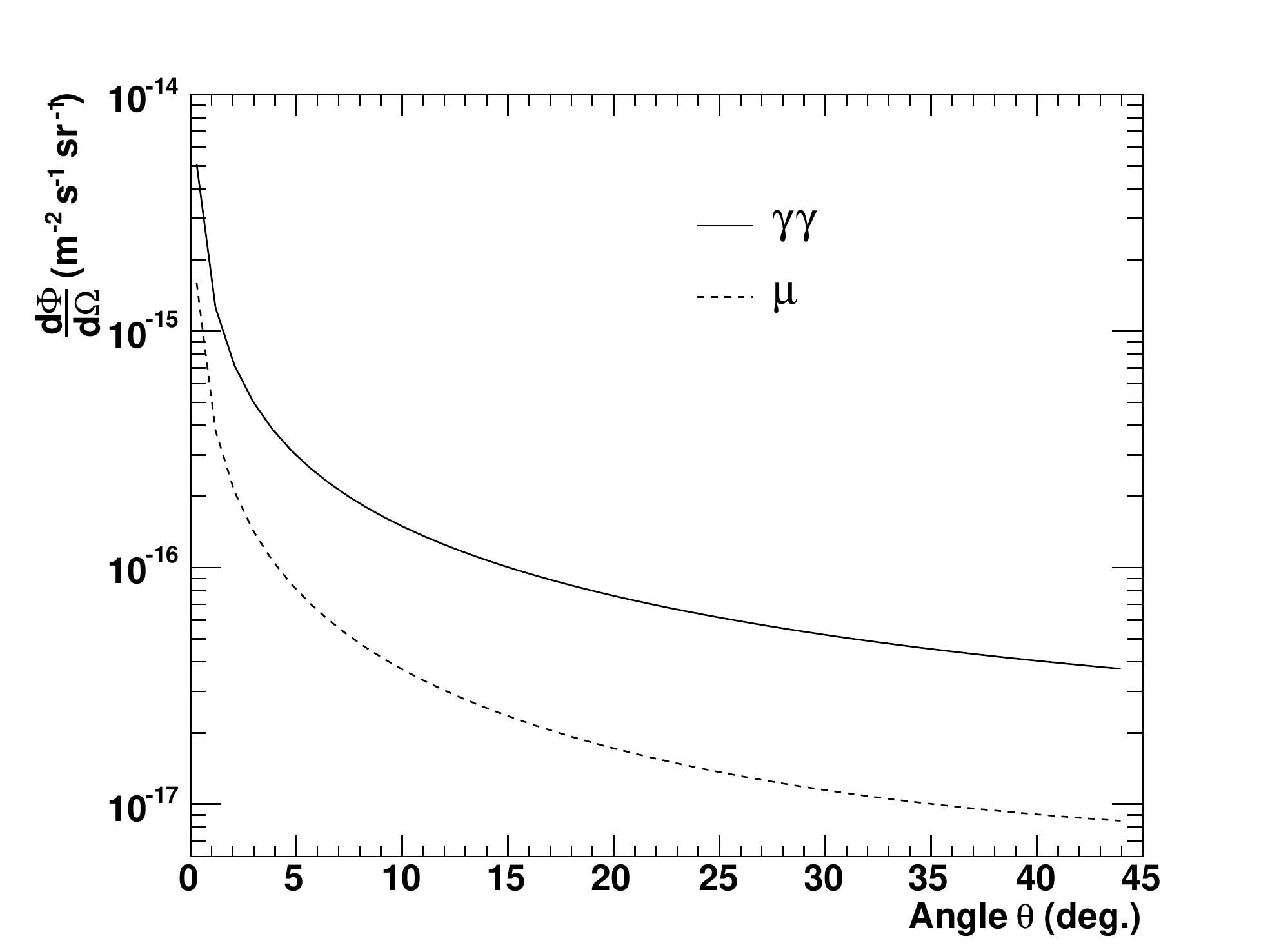}
		\caption{	\label{fig:dOmega}Differential flux of photon pairs and muons. }
\end{figure}

The flux is shown in \fref{fig:tot}, as a function of the $b_\text{max}$.  The photon curve has a positive curvature at high impact parameters because there is a non-linear relation between the impact parameter and the observation angle $\varphi$. The flux per solid angle is shown in \fref{fig:dOmega}. %It is clearly seen that t
Most of the particles come from close to the direction of the Sun. The linear rise in \fref{fig:tot} is due to the linear increase in solid angle, $\d\Omega = 2\pi b \d b$, which compensates the almost $1/b$ decrease of the differential flux $\d\Phi/\d\Omega$ seen in \fref{fig:dOmega}. In the following we consider the flux within $15^\circ$ of the Sun.


\begin{figure}[tb]
	\centering
		\includegraphics[width=\columnwidth]{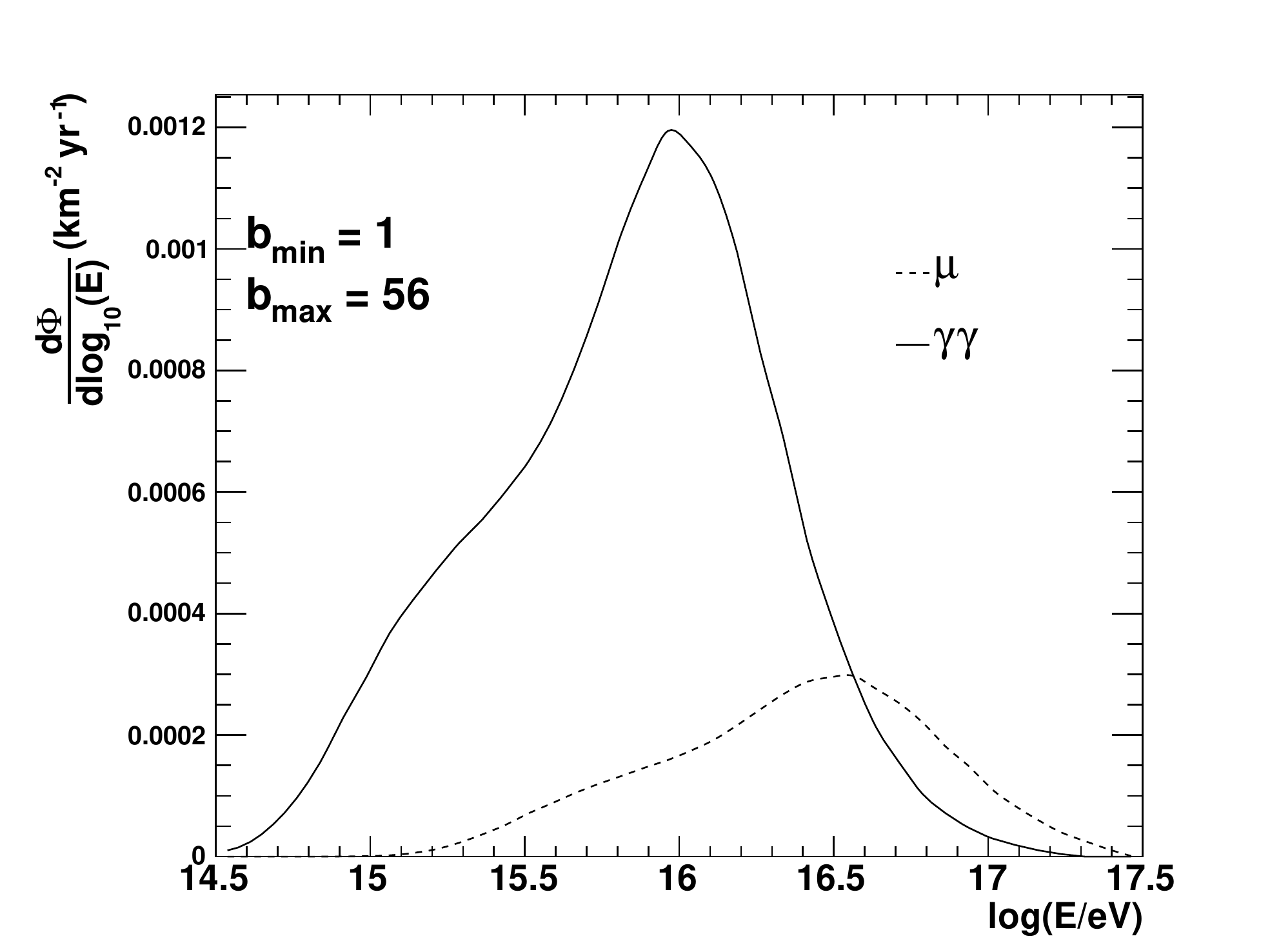}
		\caption{\label{fig:epho}Energy spectrum of muons and photon pairs (summed energy) in a $15^\circ$ area around the Sun. See the text for details.}
\label{fig:flux}
\end{figure}

The energies of the created muons and photon pairs are shown in \fref{fig:epho}. We have assumed that all interactions lead to single-pion emission, via a $\Delta^+$ with a mass fixed at $1232$ MeV and that the pions are emitted isotropiccally in the center of mass frame. The flux is integrated from $b=R_\text{Sun}$ up to $56R_\text{Sun}$ which corresponds to $15^\circ$. Since we neglect emission of two-pions, the plotted flux is lower than the actual. For photon pairs this correction is small but for muons neglecting two pion emission may lead to a significant decrease in flux. \fref{fig:tot} shows that two-pion emission accounts for close to 1/3 of the interactions that leads to muons. Hence, the muon spectrum might be low by up to 33 \%.  

\section{Detection} 

%Several detectors study highenergy cosmic rays. 
 Since muons and photons produce very different signals we will consider them separately. 

\subsection{$\mu$-signal}
The muon signal is an isolated energetic muon,  unaccompanied by a cosmic-ray shower.  The total muon flux within $15^\circ$ of the Sun is about $\SI{0.3e-3}{}$ particles/(km$^2\cdot$yr).

 Because of the limited muon lifetime, there are no other obvious sources of isolated muons.   These muons may be observed in detectors that are optimized to observe muon neutrino interactions, such as the 1 km$^3$ IceCube array.  IceCube has observed upward going muons created by neutrinos with energies up to 400 TeV \cite{PhysRevD.83.012001}.  It has also observed near-horizontal, downward-going muons with energies up to about 100 TeV \cite{Berghaus2009}.    These muons are observed at large zenith angles, so that they traverse more than 10 km of ice before reaching the detector; this ice shielding usually reduces what is initially a large bundle containing multiple muons of varying energies to a single muon.   IceCube is at the South Pole, so, over one year,  the Sun oscillates between $23^\circ$ above the horizon and $23^\circ$ below the horizon. Muons will only be visible when the Sun is above the horizon.  By selecting only very high-energy muons that point back to the Sun, the background should be very small.  Unfortunately, \fref{fig:tot} shows that Icecube is probably too small to see a signal.  Even with a larger optical Cherenkov detector like Km3Net \cite{Km3net}, the chance of observing a signal is very small. 

The next generation of neutrino detectors, like ARIANNA \cite{ARIANNA,Barwick2006} and ARA \cite{ARAprop} will have active volumes about 100 times larger than IceCube, making an observation possible.  These will detect coherent radio Cherenkov emission from neutrino induced showers, rather than muons.  For showers, their energy thresholds will be of the order of $10^{17}$ eV,  so they might be able to detect muons that radiate a significant fraction of their energy in a single electromagnetic or photonuclear interactions.    The introduction of subthreshold triggering, whereby subthreshold signals from multiple stations are combined for a clean detection \cite{GaryV}, would make these arrays into viable solar muon detectors.

ARIANNA has a planned surface area of about 900 km$^2$; at it's latitude the Sun gets up to about $30^\circ$ above the horizon.  At the midpoint of it's path, the Sun is is $15^\circ$ above the horizon, and ARIANNA will present a perpendicular area to the Sun of about 230 km$^2$; if the ARIANNA muon threshold can be reduced to of order 10 PeV, in a multi-year run, with a wide ($30^\circ$ search bin) it might be able to see a signal. 

\subsection{$\gamma$-signal} 

When a $\pi^0$ is produced near the Sun, the two photons will separate as they travel toward the Earth. 
Averaging over all photon emission angles and assuming that the photons are emitted a distance $D$ from the Earth one finds that the average photon separation is
\begin{equation}
	\langle L(\gamma_{\pi^{0}})\rangle = \frac{D\pi}{\gamma_{\pi^{0}}}.
	\label{eq:L}
\end{equation} 
The separation on Earth decreases linearly as  the $\pi^0$ energy rises.  There are a couple of ways that these photons could be detected.  At low energies, the photon separation will be too large for both photons to be visible in a single detector, and so the expected signal will be a single photon coming from the Sun which can be detected by air shower arrays.  Within a $30^\circ$ cone of the Sun, the expected total photon flux is about 0.005 yr$^{-1}$km$^{-2}$; an 200 km$^2$ array is needed to observe a signal. 

\begin{figure}[tb]
	\centering
			\includegraphics[width=\columnwidth ]{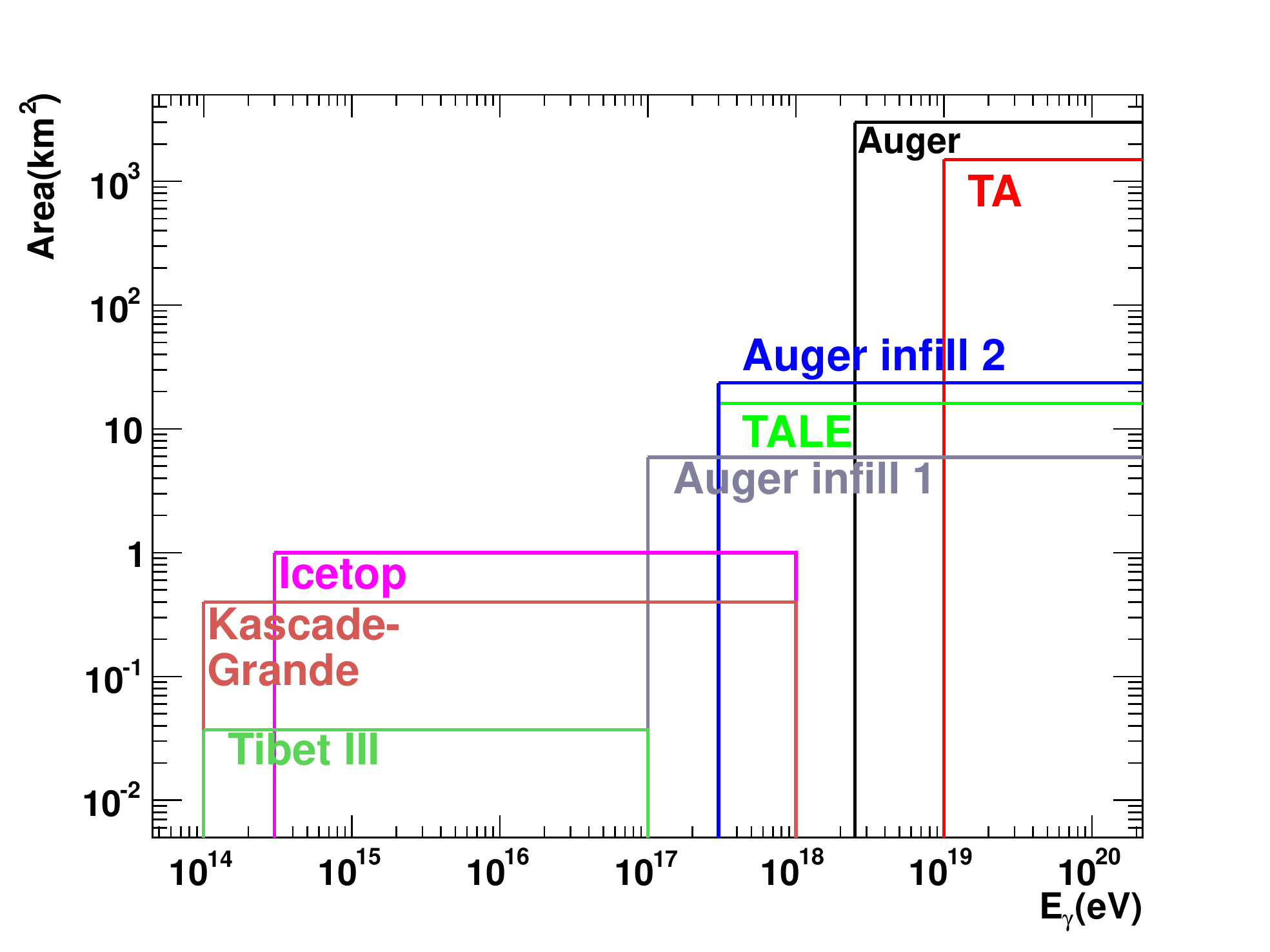}	
		\caption{\label{fig:area}(Color online) Effective area and energy thresholds for different relevant detectors \cite{Auger2010,AugerInfill,Kascade2010,TA2008,Tibet2009,Icetop2008}.}
	
\end{figure}

  \fref{fig:area} shows the energy threshold and effective area of a number of cosmic-ray detectors. The detectors IceTop \cite{Icetop2008}, KASCADE-Grande \cite{Kascade2010} and Tibet III \cite{Tibet2009} cover energies from below the knee to well above. They have surface areas around 0.1 to 1 km$^2$ and the individual detectors in the arrays are less than 150 m apart. These arrays have too small surface areas to detect a signal.  Larger detectors like the Pierre Auger Observatory \cite{Auger2010} and Telescope Array (TA) \cite{TA2008} both have the size needed to see a $\gamma$-signal. Unfortunately, both are only sensitive to particle energies above 10$^{18}$eV, which is above the bulk of the photon spectrum. The lower energy extensions of both detectors all have areas significantly smaller than 200 km$^2$ \cite{AugerInfill,TA2008} making an observation difficult. 

\subsection{$\gamma\gamma$-signal}

At high energies, a detector might observe both photons from a single $\pi^0$ giving rise to a multi-core shower.  The advantage of this channel is that the background is very low.  The probability of observing two air showers at the same time, coming from the same direction is tiny, and the probability will drop even further if one requires that they both be consistent with photon initiated showers.   An array may observe both photons if the average separation between photons is smaller than the linear dimensions of the array.   For simplicity, we assume that the arrays are square, with a typical linear dimension $\sqrt{A}$, where $A$ is the area. 

On the other hand, if the two photons are too close together, then their showers will merge into a single shower. 
For an ideal detector, this happens when the photon spacing is considerably smaller than the Moliere radius of the showers or about 70 m of air at sea level.  For real detector arrays the showers will be inseparable if the photon spacing is smaller than the typical separation between the individual detectors.  Since the largest detectors have detector spacings up to 1.5 km, this is a real limitation.

\begin{figure}[tb]
	\centering
			\includegraphics[width=\columnwidth ]{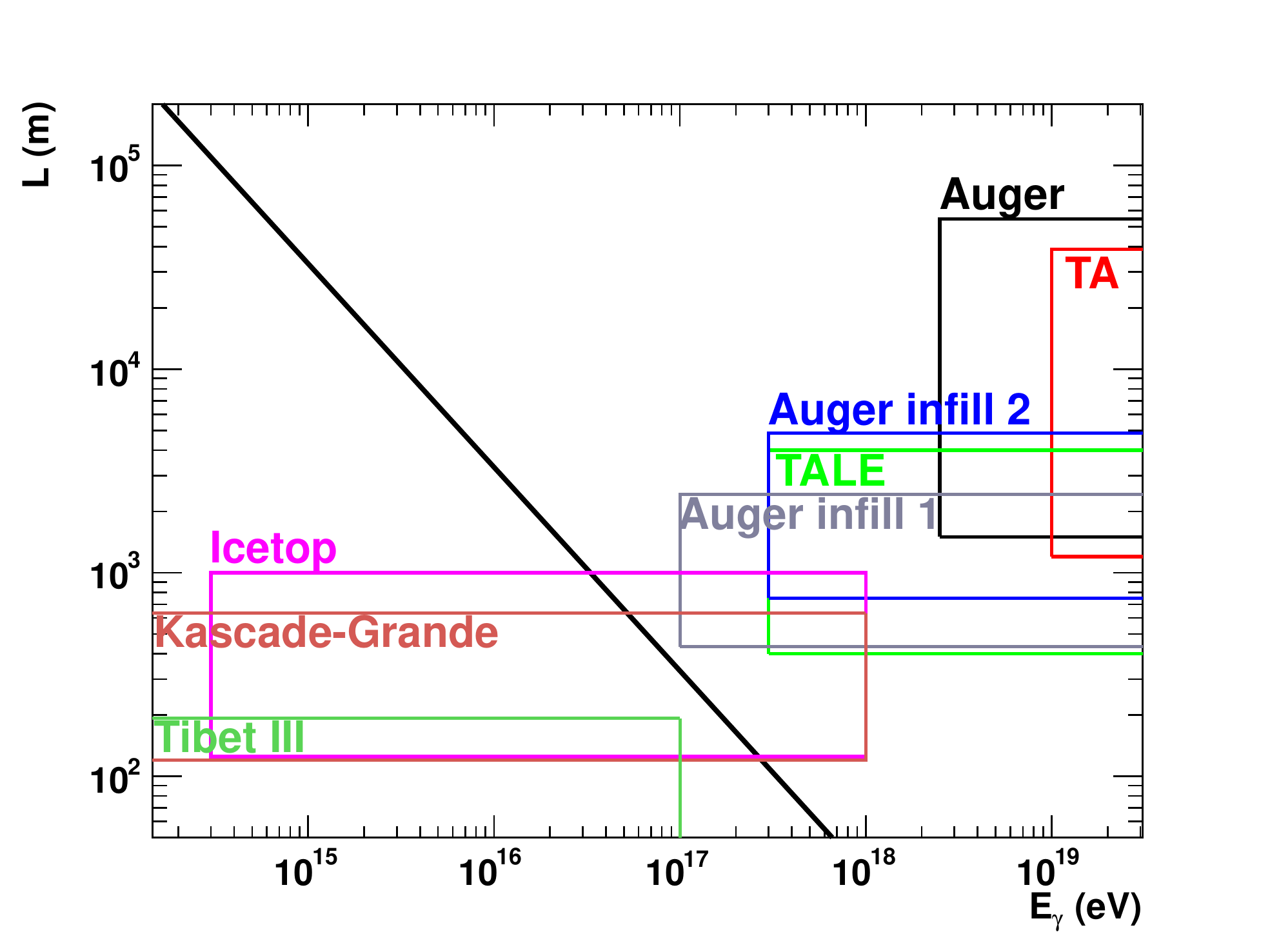}	
		\caption{	\label{fig:ldet}(Color online) Average photon separation as function of energy together with detector characteristics.}

\end{figure}

\begin{figure}[tb]
	\centering
			\includegraphics[width=\columnwidth ]{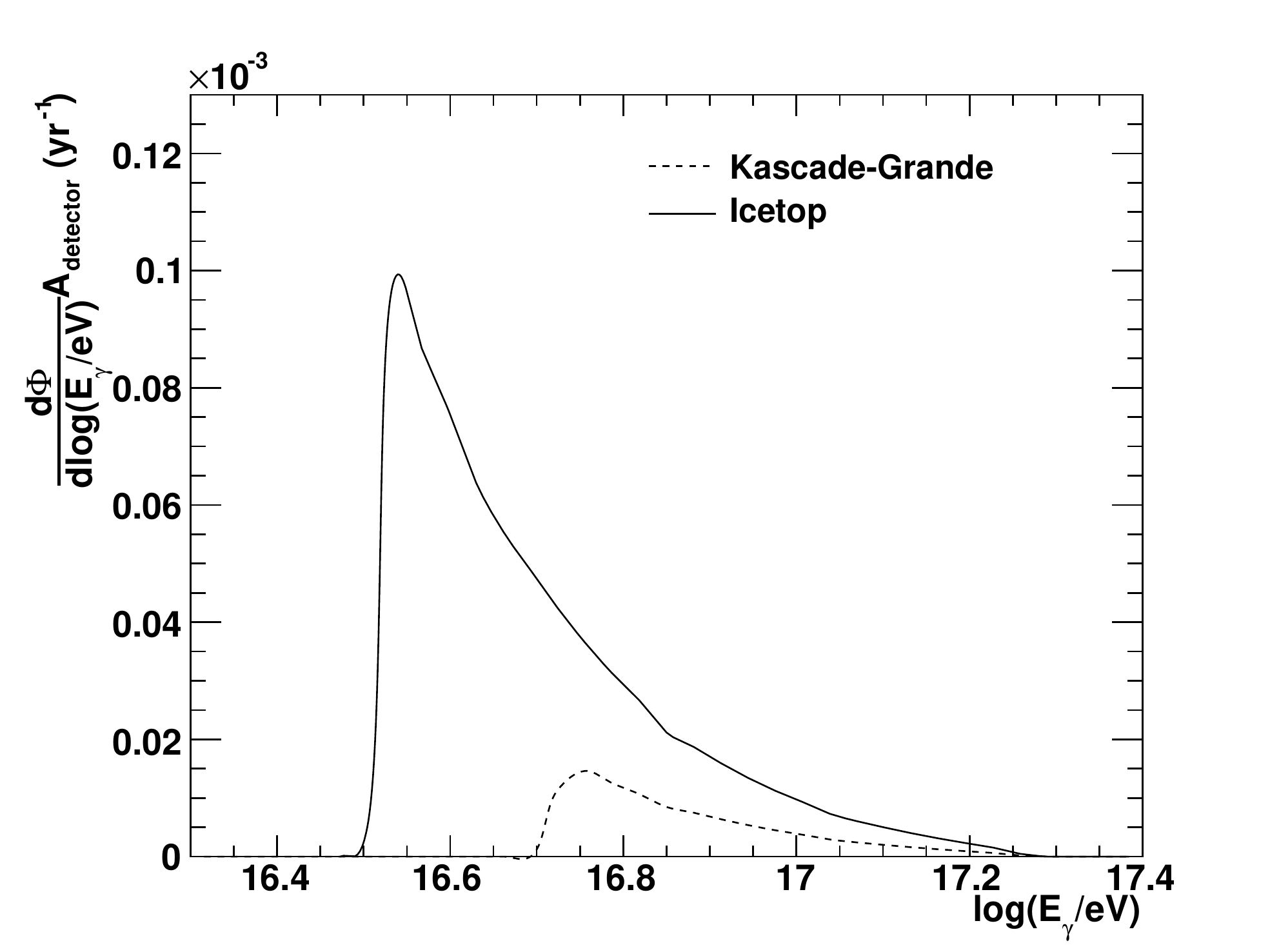}	
		\caption{	\label{fig:phodet}Photon pair flux for Kascade-Grande and Icetop. It is assumed that one observes up to $\theta=15^\circ$, only single pion emission and that the mass of the baryon resonance is 1232 MeV. }

\end{figure}

\fref{fig:ldet} compares the typical photon separations with the detection limits set by the dimensions of the detectors. Only KASCADE-Grande and IceTop have the neccessary combination of size and detector separation to be able to detect these multi-core showers.  For KASCADE-Grande and IceTop we have calculated the detected photon spectrum shown in \fref{fig:phodet}. This is done under the assumptions that single pion emission is dominant, produced via photoexcitation of a narrow resonance at an energy of 1232 MeV. The signal region is taken to be a circle around the Sun with radius  $15^\circ$. 

We assume that the detector can observe the Sun for 10 hours/day, at an average zenith angle of 45 degrees.  So, the equivalent exposure of a detector perpendicular to the Sun is about 0.3 years/year.  The expected rate is the flux times this scaling factor, multiplied by the detector area of the detector. Integrating over the spectrums one finds a multi-core shower observation rate of $N_\text{IceTop} = \SI{7e-6}{yr^{-1}}$ and $N_\text{K-Grande} = \SI{9e-7}{yr^{-1}}$.  These are both too small to be observable.

\section{Other interactions}
Similar approaches can be used to treat other interactions between cosmic rays and particles from the Sun. Here we consider two reactions: inelastic cosmic-ray solar-atmosphere interactions and $e^+e^-$ pair production. 

% Proton proton inelastisk diffraction and pair production
\subsection{Cosmic-ray interactions with the solar atmosphere}

Cosmic-ray protons may also interact with protons in the solar atmosphere, producing showers of pions.  As with $\Delta$ excitation, these pions may decay, producing photons and muons which might be visible on Earth.  Similar considerations regarding in-transit attenuation and bending apply.

We model the particle density $\rho$ of the solar corona with the Baumbach-Allen formula \cite{AschwandenCorona} which is valid up to a few solar radii. We assume that it is valid up to 5 solar radii and that the density is zero beyond this distance.  The formula gives the electron density of the corona. As long as  the corona is electrically neutral this corresponds to the proton density. We assume that the atmosphere is all protons. For a given impact parameter one can calculate the column density $\rho_c(b)$ of the corona. 

We approximate the proton-proton cross-section by a power function. From the Particle Data Book \cite[Fig. 41.11]{PDG2010} we find
\begin{equation}
	\sigma_{pp} = \SI{19.2}{mb}\times (E_{CR}/\text{GeV})^{0.0959}.
	\label{eq:sigmapp}
\end{equation} 
 We consider cosmic rays with energies between $10^{16}$ eV and $10^{19}$ eV, or, center-of-mass energies between 4 TeV and 140 TeV calculated under the assumption that the protons in the solar atmosphere are at rest. The average charged particle multiplicities $\langle N_{ch}\rangle$ are approximately \cite{JPhysG.37.083001}
\begin{eqnarray}
	\langle N_{ch}\rangle(E_{CM}) = 16.65 - 3.147\ln (E^2_{CM}/\text{GeV}^2) \notag\\
	 + 0.334 \ln^2 (E^2_{CM}/\text{GeV}^2).
	\label{eq:Nch}
\end{eqnarray}
The secondary particles will typically have a few percent of the energy of the incident cosmic ray.  We assume 50\% of the cosmic-ray energy is divided evenly between these particles (with the incident proton retaining the rest). Then, the pion energy is
\begin{equation}
	E_\pi = \frac{0.5 E_{CR}}{\langle N_{ch}\rangle(E_{CM})}.
	\label{eq:Epi}
\end{equation}
It depends only on the cosmic-ray energy since the kinetic energy of the solar protons is low compared to the cosmic rays, so $E_{CM}\approx\sqrt{2m_p^2+2m_pE_{CR}}$. From this the energies of the pions range from 87 TeV to 39 PeV.

Both neutral and charged pions are created in the interaction. Roughly 2/3 of the pions are $\pi^\pm$ and 1/3 are $\pi^0$ \cite[Tab. 41.1]{PDG2010}. The energies of the muons from $\pi^\pm$ decay are generally too low for the muons to reach the Earth. For $\pi^0$ we assume that every pion makes a shower detectable on Earth. The flux is: 
\begin{eqnarray}
	\Phi_{pp} = \frac{1}{3} \int_{10^{16}\text{eV}}^{10^{19}\text{eV}} \d E_{CR} N_{CR}(E_{CR})\times    \notag\\ 
	\langle N_{ch}\rangle(E_{CM})\times \sigma_{pp}(E_{CR}) \int \d \Omega \rho_c(b). 
	\label{eq:phipp}
\end{eqnarray} 
Differentiating \eref{eq:Epi} one can find the differential flux as
\begin{eqnarray}
	\frac{\d \Phi_{pp}}{\d E_\pi} =\frac{\d \Phi_{pp}}{\d E_{CR}}\frac{\d E_{CR}}{\d E_\pi}
	  =\frac{\d \Phi_{pp}}{\d E_{CR}} \bigg(\frac{1}{2\langle N_{ch}\rangle} -  \notag\\
	   \frac{E_{CR}m_p}{\langle N_{ch}\rangle^2 E^2_{CM}}\left[-3.147+2\cdot0.334\ln E^2_{CM}\right] \bigg)^{-1}.  
	\label{eq:dPhidEpi}
\end{eqnarray}

\begin{figure}[tb]
\centering
	\includegraphics[width=\columnwidth]{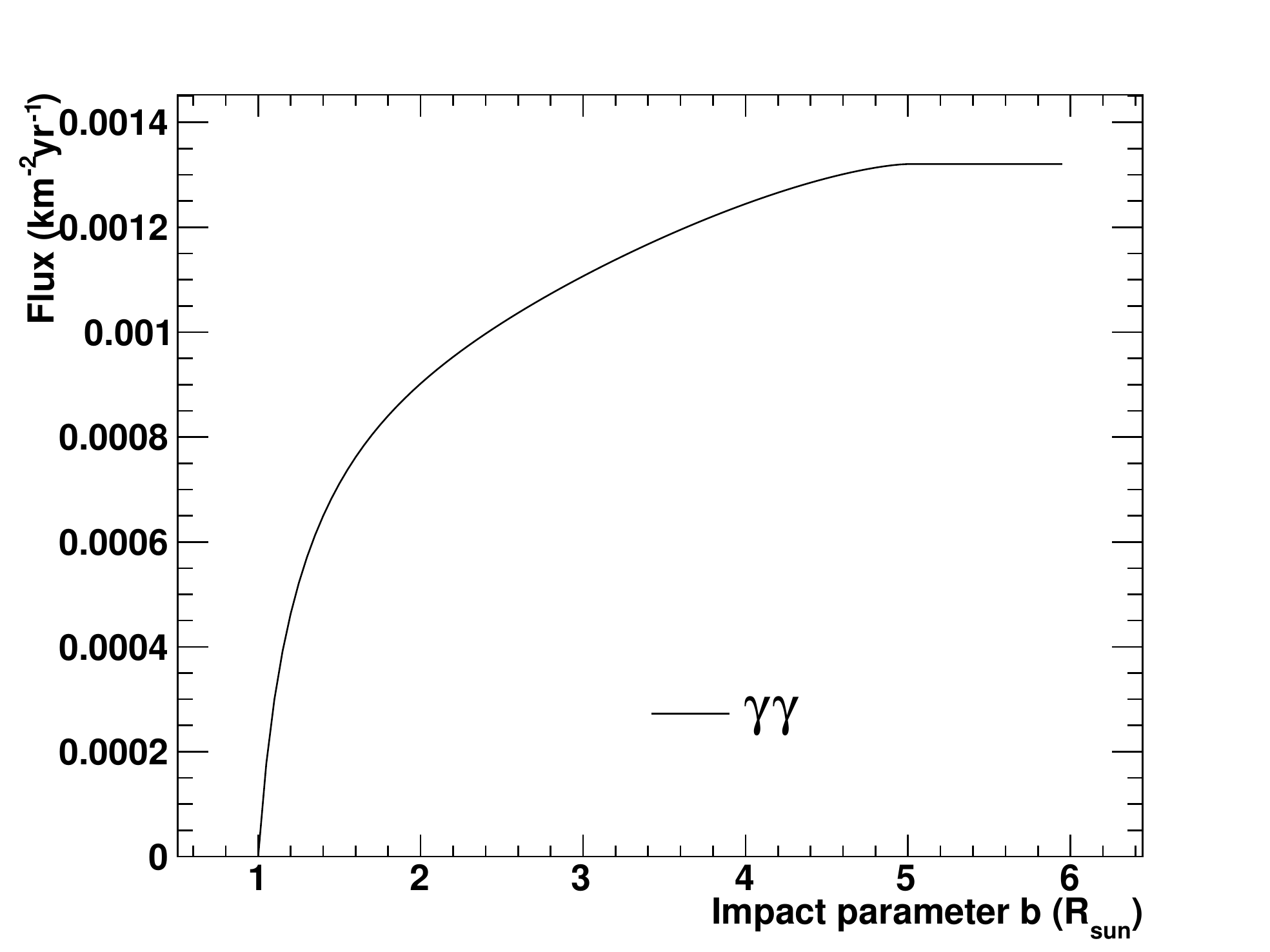}
	\caption{\label{fig:coronaflux} The flux of photon pairs produced in interactions between cosmic rays and protons in the atmosphere of the Sun coming from within an impact parameter $b$.}
\end{figure}

\fref{fig:coronaflux} shows the predicted flux as function of the maximum impact parameter in units of solar radii. The flux of photon pairs is higher than that coming from the production of $\Delta$ resonances, but the particles created in this interaction have lower energies (around $10^{14}$ eV).

It should be stressed that this estimate is very rough, with significant uncertainty regarding the region of applicability of the Baumbach-Allen formula.

%%%%%%%%%%%%%%%%%%%%%%%%%%%%%%%%%%%%%%%%%%%%%%%%%%%%%%

\subsection{Pair production}

Solar photons can produce $e^+ e^-$ pairs in the electric field of a high energy cosmic ray. The pair-production cross section is much higher than both the photoproduction cross section and the proton-proton inelastic cross section. However, the produced $e^+$ and $e^-$ have relatively low energies, and are therefore  deflected by the interplanetary magnetic field. Only the high-energy tail of the $e^+$ and $e^-$ spectrum will retain its directional information. %We assume that particles with energies above 1 TeV are only slightly deflected $(<10^\circ)$ \cite{Tibet1993}. 
Furthermore, the deflection makes the chance of reconstructing the pair very small; only a single lepton will be visible. 

The differential pair-production cross section $\d\sigma_{BH}/\d E'_+$ is given by the Bethe-Heitler formula without screening, since the cosmic rays are fully ionized \cite[Eq. 8, p. 258]{Heitler1954}. $E'_+$ is the positron energy in the frame of the cosmic ray.

The number of generated pairs from the passage of a single cosmic ray is given by
\begin{eqnarray}
	N_\text{pair}(E_{CR},b) = \int \d E'_+ \int \d t \notag\\
	 \int \d k  \frac{\d\sigma_{BH}(k')}{\d E'_{+}} \times N_\gamma(k,R(\theta)),
	\label{eq:Ppair}
\end{eqnarray}
where $k'$ is the photon energy in the rest frame of the cosmic ray. It is $k' = k\gamma(1-\cos\theta)$, where $\gamma$ is the Lorentz factor of the cosmic ray, and $\theta$ is the interaction angle defined in \fref{fig:draw}. One can conveniently define $x = (E'_+-m_e)/(k'-2m_e)$ and substitute this into \eref{eq:Ppair}. 

\begin{figure}
\centering
	\includegraphics[width=\columnwidth]{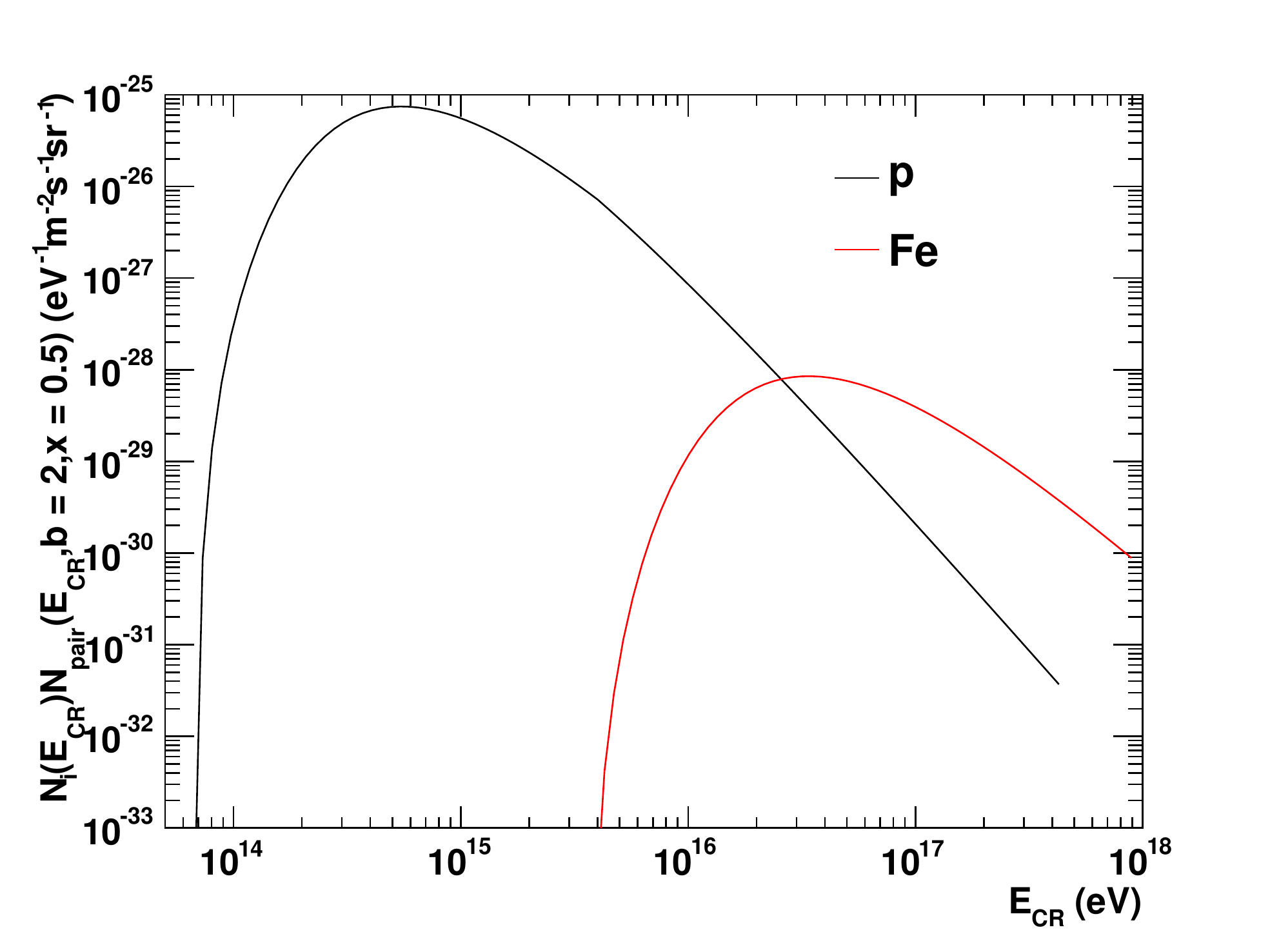}
	\caption{\label{fig:pp} (Color online) The number of generated pairs calculated by \eref{eq:Ppair} times the cosmic-ray spectrum for iron (red), $N_{Fe}$, and protons (black), $N_p$. The cosmic rays travel at impact parameter $b=2$.}
\end{figure}

\fref{fig:pp} shows the number of pair productions per cosmic ray times the cosmic-ray spectrum. For the proton spectrum we use \eref{eq:Ncr} and 
\begin{equation}
N_p(E<E_\text{knee})=\SI{7.3e19}{\frac{eV^{1\pnt7}}{m^{2}.s^{1}.sr^{1}}} E^{-2.7},
	\label{eq:Npknee}
\end{equation}
for energies below the knee; $E_\text{knee} = \SI{4e15}{eV}$. The iron spectrum is less certain. Results from KASKADE-Grande \cite[Fig. 4]{AstroparPhys.31.86} indicate a flux of 
 \begin{equation}
N_{Fe}(E)=\SI{3.2e15}{\frac{eV^{1\pnt5}}{m^{2}.s^{1}.sr^{1}}}E^{-2.5},	
	\label{eq:NFe}
\end{equation}
for energies from $10^{16}$ eV to $10^{17}$ eV. We assume that it is valid up to $10^{18}$ eV.  

By integrating \eref{eq:Ppair} over cosmic-ray energy and impact parameter one can get an impression of this flux. Integrating up to a $15^\circ$ angular separation one finds a flux around 100 $e^+$ and $e^-$ per sq. km. per year. Simple kinematic calculations indicate that almost all $e^+$ and $e^-$ have energies that are lower than 1 TeV and hence will be deflected substantially by the interplanetary magnetic field \cite{Tibet1993}. Hence, the low flux and the absence of a directional variation make this contribution to the cosmic electron-positron flux almost impossible to detect.

\section{Conclusion}
% Conclusion
We have shown that photoproduction of $\Delta^+$ resonances around the Sun create high-energy pions that decay to either muons or photons. Observing a high-energy muon or photon pair from the Sun is a clear signature of such an interaction giving a good probe of the total cosmic nucleon flux. With current detectors the chance of observing this interaction is small but may be within reach of future detectors.

\bibliography{biblio}	

\end{document}